\documentclass[prl,amsmath,amssymb,reprint]{revtex4-1}

\usepackage{blindtext}
\usepackage{amsmath}
\usepackage[capitalise]{cleveref}
\usepackage{siunitx}
\usepackage{graphicx}

\newcommand{\mb}[1]{\mathbf{#1}}

\begin{document}
\title{Efficient global structure optimization with a machine learned surrogate model}
	
\author{Malthe K. Bisbo}
\author{Bj{\o}rk Hammer}

\affiliation{
Department of Physics and Astronomy, Aarhus university, DK-8000 Aarhus C, Denmark		
}

\date{\today}

\begin{abstract}
We propose a scheme for global
optimization with \textit{first-principles} energy expressions (GOFEE) of atomistic structure.
While unfolding its search, the method actively learns a
surrogate model of the potential energy landscape on which it performs
a number of local relaxations (exploitation) and further structural
searches (exploration). Assuming Gaussian Processes, an acquisition
function is used to decide on which of the resulting structures is the
more promising. Subsequently, a single point
\textit{first-principles} energy calculation is conducted for that
structure. The method is demonstrated to outperform by two orders of magnitude
a well established \textit{first-principles} based evolutionary algorithm in finding
surface reconstructions. Finally, GOFEE is utilized to identify initial stages of the
edge oxidation and oxygen intercalation of graphene sheets on the Ir(111) surface.
\end{abstract}

\maketitle

In materials science and physical chemistry, the search for optimal
structure is a recurring task, e.g.\ in describing crystalline
defects, such as grain boundaries \cite{grain_bnd:zande} and surface
reconstructions \cite{reconstruction:TiO2, SnO2}, and in modeling heterogeneous systems such as
binary compounds \cite{binary:SnO2,binary:review} and supported nano-particles
\cite{clusters:alexandrova1,clusters:alexandrova2,LEA}.  Depending on the search strategy and
the complexity of a given problem, many thousands of
energy and force evaluations may be required for the structural
candidates in the course of the search. The results of these
calculations constitute a set of structure-energy relation data points
which represents a valuable resource that can direct the search. If
the energy calculations are done at a \textit{first-principles} (FP) level
using density functional theory or quantum chemical methods, the
computational bottle-neck lies in performing the individual
energy/force evaluations and considerable speed-ups may be achieved by
introducing machine-learning techniques that utilize this resource and
provide tools to minimize the total amount of FP calculations.

As an example, a machine-learning technique may be utilized to estimate, based on the data set, the potential energy of each atom in a given structure at practically no computational cost \cite{thomas_prl}. This information may then be used to guide further structure search, e.g.\ by shifting or replacing the least stable atoms, resulting in new structures that are more relevant to the search.

Greater is the potential for approaches where the entire search, including local relaxations, is performed on a machine learned surrogate energy landscape. As a first approach, the ML model can be trained in advance on a database of structures and FP properties.
Here, regression models based on kernels, invariant polynomials and neural networks have all proven successful in a number of studies \cite{parrinello_behler,GAP,MTP,schnet} and are drastically changing the field of fitting force fields \cite{NNsearch:alexandrova,FF:roitberg,FF:ramprasad,FF:GDML,GAP:carbon}. 

Since the models are all interpolative and give reliable results only within their training domain, pre-fitted models have the drawback that they require the expensive generation of a large, diverse database of training data to be successfully applied to a structure search problem.

A more data efficient approach is to start from a small incomplete training database and then augment it on the fly only with the data deemed most relevant, at that time, for solving the task at hand e.g.\ structure optimization \cite{SS:alexandrova,active:rinke,active:oguchi,GAPRSS:boron,GAPRSS:crystal,activeSS:calypso,activeSS:deringer,activeSS:shapeev,SS_dftb:maxime}. This is the philosophy in the area of \textit{active learning} \cite{activeFF:roitberg,activeFF:Zhang}. It was recently demonstrated in the context of an evolutionary algorithm (EA) structure search framework, where an artificial neural network was trained and used for local relaxation while the EA acted only on FP single-point energy evaluations \cite{NN_SS:ouyang,LEA}.
Active learning approaches have been extensively applied in molecular dynamics simulations \cite{activeMD:Vita,activeMD:andrew,activeMD:Ohno,activeMD:evgeny,activeMD:kresse} with data efficient training databases as a byproduct. It has also been applied in \textit{local} optimization problems such as local relaxation \cite{local:karsten, local:kastner} and in minimum energy path determination with the Nudged Elastic Band (NEB) method \cite{neb:andrew2016, neb:hannes2017, neb:bligaard2019}. Local optimization problems lend themselves particularly well to the construction of surrogate energy landscapes as a Cartesian coordinate representation of atoms may be adopted.

When a surrogate energy landscape is trained via active learning, the issue
arises which next computationally expensive FP single point energy to evaluate. In this work, we present a strategy, GOFEE, that utilizes Bayesian statistics in the context of Gaussian Processes (GP) \cite{GP:Rasmussen}.  This framework allows for the estimation of the uncertainty in any prediction on the surrogate energy landscape, which provides the foundation for an acquisition function that guides the search.  The method is first demonstrated to work up to two orders of magnitude faster for the identification of reconstructed surfaces of rutile $\text{SnO}_2(110)\text{-}(4\times1)$ and anatase $\text{TiO}_2(001)\text{-}(1\times4)$. Next, the method is used to tackle a set of otherwise prohibitively complex problems for graphene patches on Ir(111): the edge structure itself, the structure of the oxidized edges, and the pathways for oxygen intercalation at the graphene edges.

\begin{figure}
	\centering
	\includegraphics[width=0.9\linewidth]{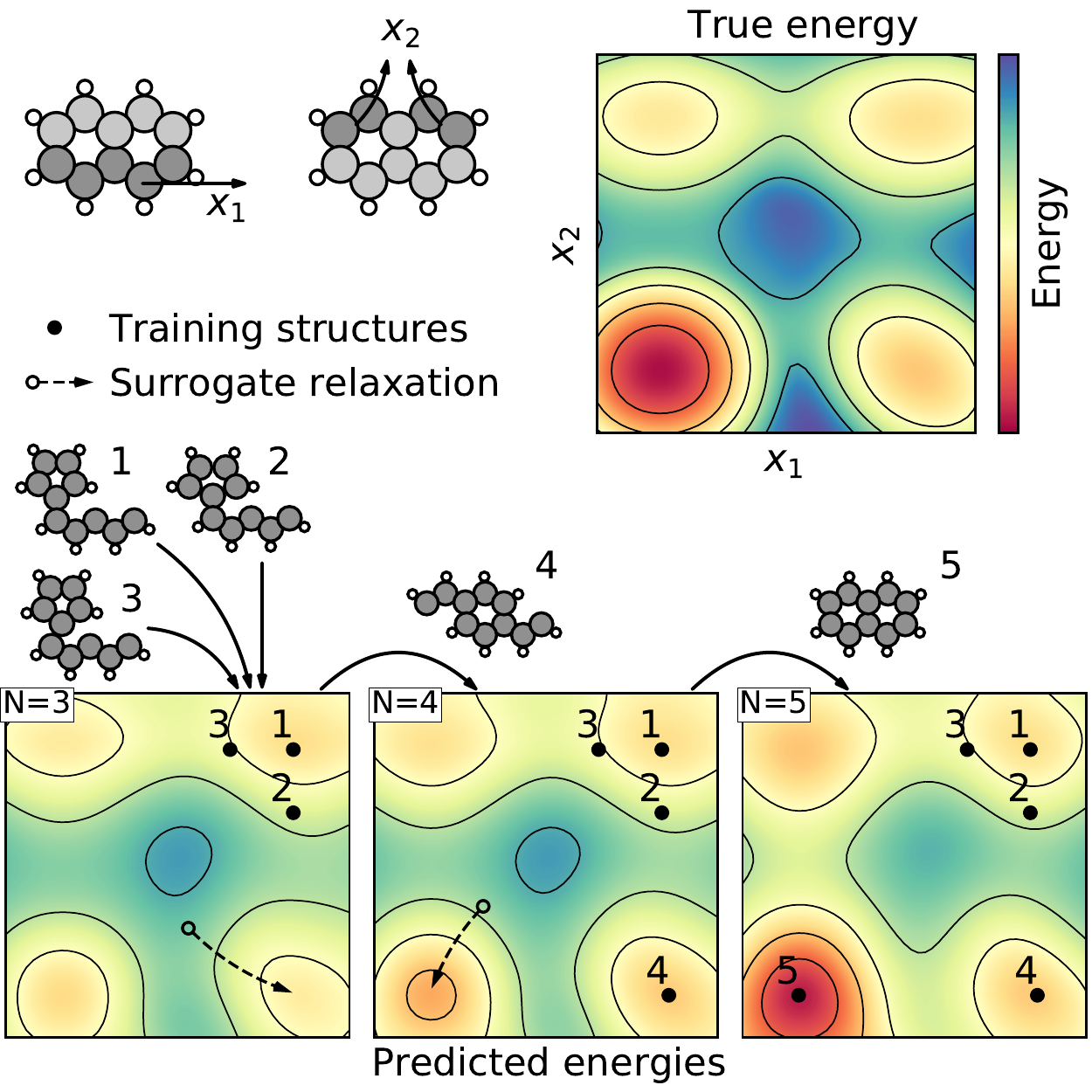}
	\caption{Example of a surrogate guided structure search in a two dimensional search space. The two artificial dimensions are constructed by perturbing naphthalene as depicted in the top left. The FP energy landscape is shown in the top right. In the bottom, the surrogate landscape resulting from $N=3$, 4 and 5 training structures is shown. Note that training structures 4 and 5 result from local relaxation in the surrogate landscape.}
	\label{fig:naphthalene}
\end{figure}


The degree of success achievable by any surrogate based search method is largely dependent on the quality of the surrogate model \cite{GAPRSS:boron,GAPRSS:crystal}. In this work, the GP regression method is adopted partly because of its tractable simplicity and partly because GPs are expected to behave well, as the number of training examples increase during the search, due to the adaptive predictive power inherent to non-parametric methods.


The task of GPs is to infer a distribution over functions, here $p(E_{sur}|X,\mb{E})$, that is consistent with a training set of observed atomic configurations $X=(\mb{x}_1,\mb{x}_2,\dots,\mb{x}_N)^T$ and their corresponding energies $\mb{E}=(E_1,E_2,\dots,E_N)^T$. To include the symmetries of the system, $\mb{x}_i$ is taken to be the feature vector for the i'th configuration rather than the Cartesian coordinates. We adopt the global fingerprint feature from Oganov and Valle \cite{oganov_valle}, however the method is expected to work equally well with other features. For GPs, the distribution is assumed to be normal, which enables estimation of not only the energy $E_{sur}(\mb{x})$ as the mean of the distribution, but also the predictive uncertainty $\sigma_{sur}(\mb{x})$. As will be discussed later, the predictive uncertainty is useful in a search context, as it allows the distinction between explored and unexplored regions of the search space.

A Gaussian process is specified by it's prior mean function $\mu(\mb{x})$ and covariance function $k(\mb{x}_i,\mb{x}_j)$, which encodes prior assumptions about the target function. Given these, energy and uncertainty predictions for a new structure $\mb{x}_*$ is carried out using
\begin{align}
E_{sur}(\mb{x_*}) &= \mb{k}_*^T(K+\sigma_n^2 I)^{-1}(\mb{E}-\mu(\mb{x}))+\mu(\mb{x}), \\[4pt]
\sigma_{sur}(\mb{x_*})^2 &= k(\mb{x}_*,\mb{x}_*) - \mb{k}_*^T(K+\sigma_n^2 I)^{-1}\mb{k}_*,
\end{align}
where $K = k(X,X)$ and $\mb{k}_* = k(X,\mb{x}_*)$ and the target function is assumed noisy with uncertainty $\sigma_n^2=10^{-5} \SI{}{\electronvolt}^2$, which acts as regularization. To include the repulsive atomic core generally present, the prior mean function is taken to be a conservatively chosen repulsive interatomic potential, specifically $\mu(\mb{x})\propto\sum_{ij}(0.7*r_{CD,ij}/r_{ij})^{12}$, where $r_{ij}$ and $r_{CD,ij}$ are the distance and covalent distance between atom $i$ and $j$. This is especially beneficial in a structure search context, where the fine details of the repulsive part of the potential is not crucial, unlike the near equilibrium part of the potential.
The covariance function was chosen to be a sum of two Gaussian covariances
\begin{align}
k(\mb{x},\mb{x}') = \theta_1\exp\left(-\frac{(\mb{x}-\mb{x}')^2}{2l_1^2}\right) + \theta_2\exp\left(-\frac{(\mb{x}-\mb{x}')^2}{2l_2^2}\right)
\end{align}
one with a large characteristic length scale $l_1$ carrying most of the weight $\theta_1=(1-\beta)\theta_0$, and one with a smaller length scale $l_2$ and less weight $\theta_1=\beta\theta_0$, where $\beta=0.01$ and $\theta_0=\mb{y}^T(K+\sigma_n^2 I)^{-1}/N$. In the $\text{SnO}_2$ and $\text{TiO}_2$ application we use $l_1=10$ and $l_2=2$ whereas for the graphene edge $l_1=10$ and $l_2=0.5$ were used.
This enhances the models capacity to capture both the long scale features of the energy such as large energy funnels, as well as allowing for sufficient resolution of local minima.


\Cref{fig:naphthalene} shows, for a simple problem how the surrogate energy landscape improves as more data is added to the training set. The system considered is naphthalene, constrained, for illustrative purposes, to change only according to the two coordinates specified. The resulting 2d slice of the full energy landscape, contains four local minima including naphthalene itself.
With only a few training examples near one local minimum, the model is able to predict the locations of the remaining minima to approximately coincide with those of the true energy landscape. In the search we will take advantage of this, and conduct most of the search, specifically all local relaxations, in the surrogate energy landscape, which is orders of magnitude faster than FP calculations. As illustrated in the figure, a structure relaxed with the model can then be evaluated with a single FP calculation and used to update the model.

\begin{figure}
	\centering
	\includegraphics[width=0.9\linewidth]{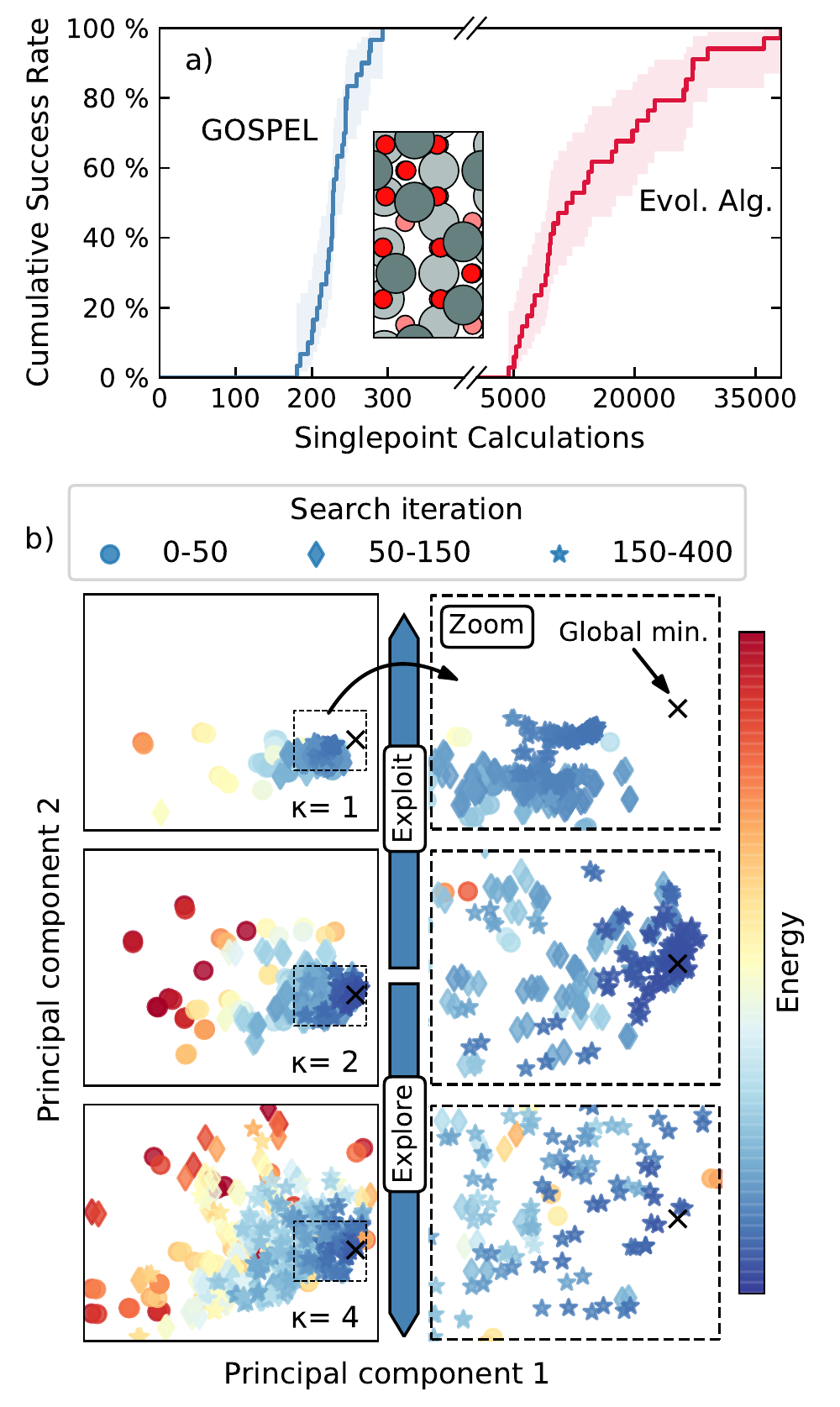}
	\caption{Success curves for finding the global minimum of a) $\text{SnO}_2(110)\text{-}(4\times1)$ when using GOFEE compared to a well established evolutionary algorithm. The global minimum structure is shown as an inset.
	In c) the structures evaluated during searches with $\kappa=1,2$ and $4$ respectively are depicted in 2d using PCA on the feature space. A suitable value for $\kappa$ strikes a balance between over-exploitation ($\kappa=1$) and over-exploration ($\kappa=4$) in the search.}
	\label{fig:pcaplot}
\end{figure}

Relying entirely on the surrogate model to guide the search has the drawback that the data collection process, vital to actively improving the model, is itself model dependent. This interplay has a tendency to cause under-exploration of the search space in turn leading to premature stagnation of the search.
The minimum belonging to naphthalene in \cref{fig:naphthalene} is an example that the true depth of a minimum might be underestimated until appropriate data has been collected. To remedy this problem we bias data collection towards unexplored regions of the search space, using the predictive uncertainty $\sigma_{sur}(\mb{x})$ as a natural way to quantify this.
Data collection is then performed according to an acquisition function $f(x)$, relying on both the predicted energy and uncertainty. There exist multiple choices for such an acquisition function \cite{acquisition}, \textit{expected improvement} and \textit{probability of improvement} are some, as well as the \textit{lower confidence bound}
\begin{align}
f(\mb{x})=E_{sur}(\mb{x}) - \kappa \cdot \sigma_{sur}(\mb{x}) ,
\label{eq:aquisition}
\end{align}
used in this work due to its simplicity. Here $\kappa$ is a tunable parameter determining the emphasis on the predicted uncertainty and thus the degree of exploration in the search.

The surrogate model is central to the GOFEE search method, which is initialized with a small set of randomly generated structures, for which the FP energy is evaluated. They make up the first structures in a database used for training the surrogate model, and to which all subsequent FP evaluated structures are added.

A diagrammatic layout of GOFEE is given in Fig. S1.
Training the surrogate model is the first step in each search iteration after which a diversified population of the best structures, currently found in the search, is used to generate a set of new candidate structures using simple rattle mutations as in Monte Carlo and evolutionary search strategies. 
To take full advantage of the computational inexpensiveness of the surrogate model, multiple new candidates are generated and relaxed in each search iteration instead of just one as shown for the example in \cref{fig:naphthalene}. From all these relaxed candidates only the single most promising, as estimated by the acquisition function, is evaluated with FP. To accommodate some force information without training on forces directly, this structure is perturbed slightly in the direction of the force and a second FP calculation is performed on the resulting structure. A search iteration is concluded by adding these two structures to the training database.

As a first example using GOFEE we considered the $\text{SnO}_2(110)\text{-}(4\times1)$ surface, for which the global minimum structure is known \cite{SnO2} and is shown as an inset in \cref{fig:pcaplot}a. The figure also shows the cumulative success curves for finding the global minimum with this method as well as with the well established EA originally used to find the structure \cite{GA}. 
Noting the broken axis, the figure shows a two orders of magnitude decrease in the number of FP calculations required to reach, e.g., $80\%$ success. This is largely attributed to the fact that this method relies only on FP for single point calculations.
To show the effect of the exploration promoting parameter $\kappa$, a specific search instance for each of $\kappa=1,2,4$ is shown in \cref{fig:pcaplot}c. In the figure, principal component analysis (PCA) is used to project all structures visited in each of the three searches onto the same two principal components determined from a large set of structures. 
The chosen search instances showcase common behavior for the three values of $\kappa$. For $\kappa=1$ the search has a tendency to over-exploit local minima, and as a result get stuck in a local minimum before reaching the global minimum. 
For $\kappa=4$ the opposite is true and the search will superficially explore many local minima before starting to optimize the best of these. The $\kappa=2$ search represents the optimal compromise between the two, performing a necessary but sufficient amount of exploration before settling to optimizing the global minimum. In all three examples it is apparent that high energy structures are primarily sampled in the beginning of the search, when the surrogate model is still learning the rough features of the energy landscape.
GOFEE was similarly applied to the $\text{TiO}_2(001)\text{-}(1\times4)$ surface reconstruction problem, displaying the same degree of improvement as compared to the EA. The results are shown in Fig. S2.  


\begin{figure}
	\centering
	\includegraphics[width=1.0\linewidth]{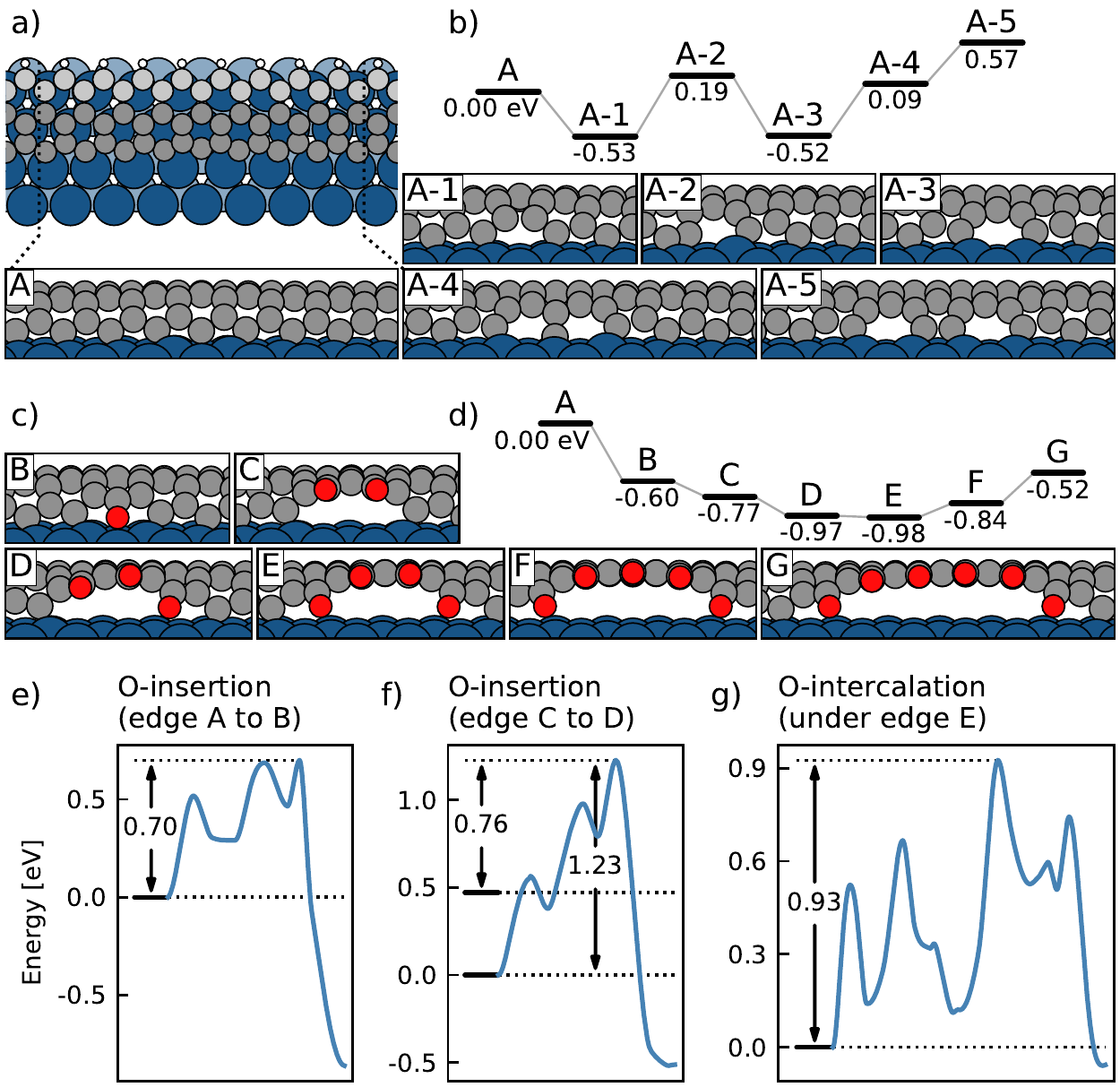}
	\caption{a-b) The most stable structures from the search and their energies shown for the edges with 0-5 carbon atoms less than the perfect edge. c-d) Structures and energies for oxygen added to the perfect edge. Structures (E-G) are made by hand, motivated by the trend from structures (B-D). Finally lowest energy CI-EB curves are given for: e) Inserting a surface adsorbed oxygen atom to the perfect edge (A), f) inserting the third oxygen atom and g) intercalation of an oxygen atom below the edge of structure (E). In the search, the second layer of iridium is kept fixed, while the upper layer is allowed to relax a maximum of half a covalent distance away from the bulk positions.}
	\label{fig:graphene_edge}
\end{figure}

To demonstrate the versatility of the GOFEE method we proceed to address the hitherto prohibitively complex problems of determining the edge structure of graphene patches on Ir(111) as well as that of the oxidized edge.
The resulting structures are used to study the atomistic mechanisms involved in intercalation of oxygen in the system.
The intercalation process has been intensively studied experimentally for Ir(111) \cite{gr:Michely2012,gr:Michely2016,gr:liv2012} and involves dissociative adsorption and diffusion of oxygen as well as penetration of the graphene edge. Although experiments suggest \cite{gr:Michely2012} that the limiting step for the intercalation process is this edge penetration, it is not well understood.

In our contribution to fill this gap, intercalation through the non oxidized edge was first considered. \Cref{fig:graphene_edge}a shows the most stable edge structures found, using our search method, when varying the number of carbon atoms present in the cell. Using the energy of carbon within the graphene patch as reference, the energies are compared in \cref{fig:graphene_edge}b, showing that the preferred structure is not the perfect edge (A), but instead the structures with one (A-1) and three (A-3) carbon atoms less on the edge, both of which feature pentagonal rings (see Fig. S3). This can be attributed to the fact that these structures avoid forming the unfavorable carbon iridium bond in the position of largest mismatch between the periodicities of the graphene edge and the iridium surface.
Although this effect does cause small gaps in the graphene edge, it is not enough to allow for oxygen intercalation, with the calculated energy barriers being larger than $\SI{2}{\electronvolt}$ for all structures.

The structure of the oxidized graphene edge was also considered, as oxygen is naturally present during the intercalation process. Searches are performed with up to three oxygen atoms in the cell. The resulting structures and energies are depicted in \cref{fig:graphene_edge}c (B-D), Fig. S4 and \cref{fig:graphene_edge}d. They display a preference for oxidizing the edge with the oxidized region partially detaching from the surface when two or more oxygen atoms are present. This results in a significant gap in the edge likely of accommodating intercalation. \Cref{fig:graphene_edge}c (E-G) (Fig. S4) further shows the structures resulting from extending this trend up to six oxygen atoms. For the energies, atomic oxygen adsorbed on the iridium surface is used as the reference. Based on the energies, the size of the gaps are thermodynamically self limiting, as edge oxidation is only thermodynamically favored up to four oxygen atoms. Further oxidation requires breaking of increasingly strong C-Ir bonds.

To study whether these oxidized edge structures are likely to form and contribute to the intercalation process under typical experimental conditions, lowest energy paths were calculated using the climbing image elastic band (CI-EB) NEB type method \cite{autoNEB}. \Cref{fig:graphene_edge}e and f, (Fig. S5-6) respectively shows the energy profiles for inserting the first and third oxygen to the edge revealing the third oxygen to be the more expensive of the two with an energy barrier of $\SI{1.23}{\electronvolt}$.
However oxygen intercalation experiments typically feature large oxygen coverages on the iridium surface, resulting in weaker bonding of the adsorbed oxygen as this coverage builds up. This effectively lowers the barriers, as the transition state for binding to and opening the graphene edge is expected to remain unchanged. The effect is depicted in \cref{fig:graphene_edge}f showing how the initial state energy, of the oxygen being inserted, is increased when sharing a single iridium atom with a neighboring adsorbed oxygen atom.
\Cref{fig:graphene_edge}g (Fig. S7) shows the energy profile for the intercalation of an oxygen atom through the edge gap of structure (E), displaying a barrier of $\SI{0.93}{\electronvolt}$, which will also be lower at realistic oxygen coverages as discussed above.

In conclusion we have formulated a machine learning enhanced structure search method and 
used it to solve a previously prohibitively hard problem. This has provided insight into the atomistic structure of graphene island edges involving pentagonal rings and into the atomistic mechanisms of oxygen intercalation for graphene on Ir(111).

\bibliographystyle{apsrev4-1}
\bibliography{references}

\pagebreak

\widetext
\begin{center}
	\textbf{\large Supplemental Materials: Efficient global structure optimization with a machine learned surrogate model}
\end{center}
\setcounter{equation}{0}
\setcounter{figure}{0}
\setcounter{table}{0}
\setcounter{page}{1}
\makeatletter
\renewcommand{\theequation}{S\arabic{equation}}
\renewcommand{\thefigure}{S\arabic{figure}}
\renewcommand{\bibnumfmt}[1]{[S#1]}
\renewcommand{\citenumfont}[1]{S#1}

\section*{GOFEE search method}
\begin{figure}[b!]
	\centering
	\includegraphics[width=0.6\linewidth]{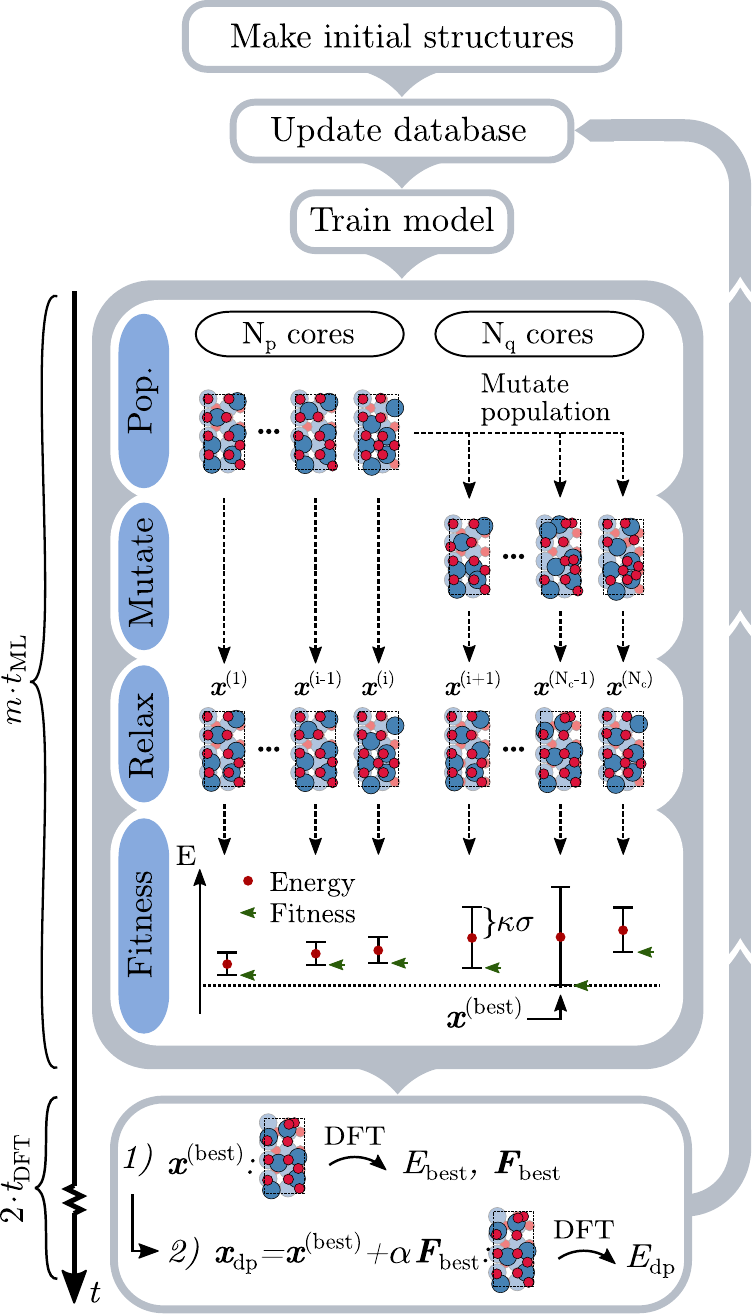}
	\caption{Sketch of the GOFEE search method depicting the key elements in the search method. i) Creation and first principles evaluation of random initial structures. ii) Addition of new first principles data to training database and training of Gaussian Process regression model. iii) Generation of a set of new candidate structures by applying mutation operations to the population. iv) Local relaxation of the population and new candidates with the model. v) Selection of the most promising relaxed candidate using the \textit{lower confidence bound} acquisition function. vi) Single point first principles evaluation of the chosen structure and another on this structure perturbed slightly along the forces. Finally the search is carried out by repeating step ii)-vi).}
	\label{fig:method}
\end{figure}

A sketch of the GOFEE search method is shown in \cref{fig:method} and depicts the key elements in the search. 

\section*{$\text{TiO}_2$ benchmark and scaling}
\begin{figure}[b!]
	\centering
	\includegraphics[width=0.6\linewidth]{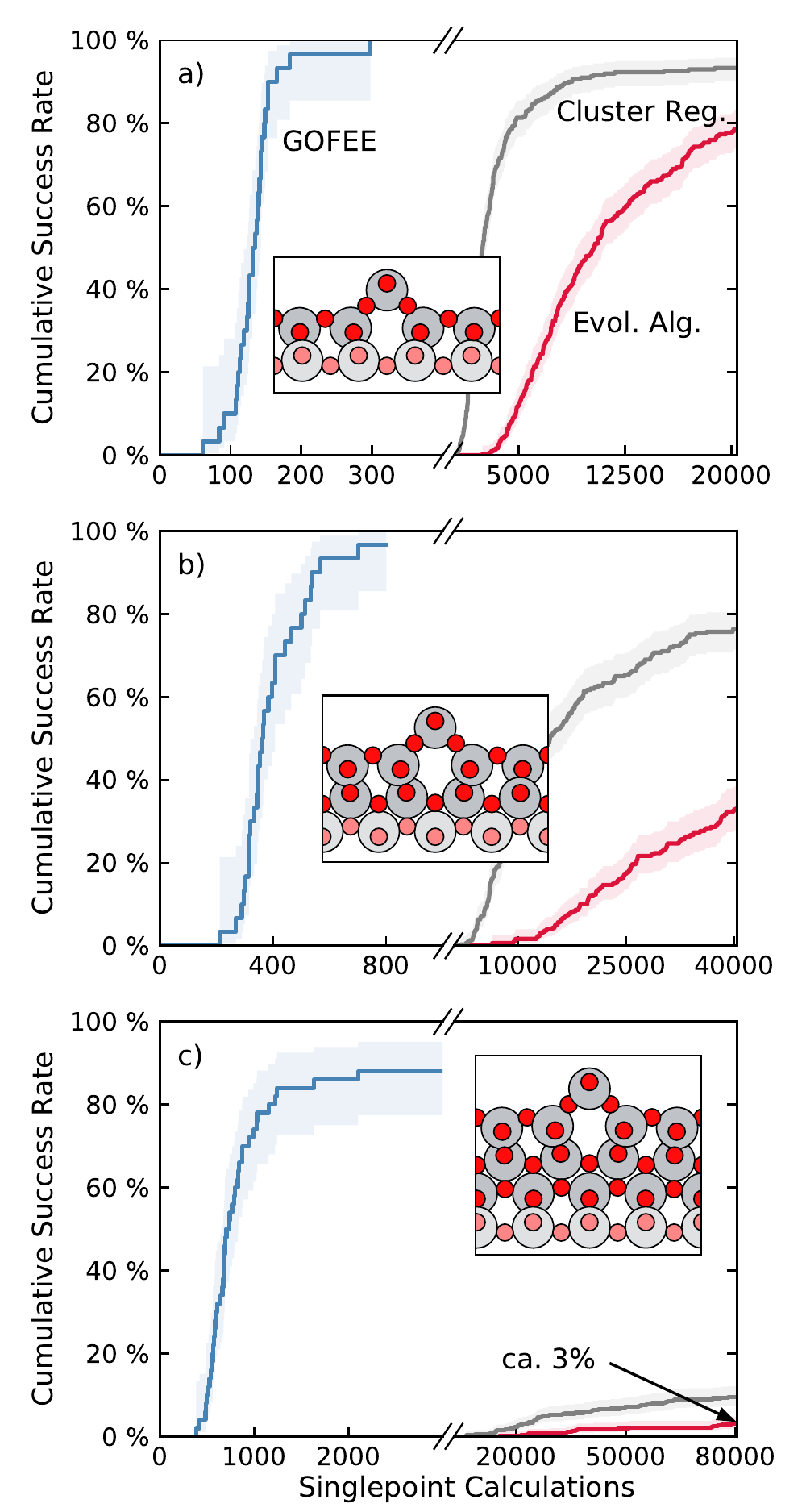}
	\caption{Comparison between the three structure search algorithms GOFEE, a well established evolutionary algorithm and this evolutionary algorithm with only the candidate generation step enhanced by machine learning. a)-c) Show success curves (note the broken first axis) comparing these three structure search algorithms on the anatase $\text{TiO}_2(001)\text{-}(1\times4)$ surface reconstruction with the first layer fixed to the bulk positions. In the three plots, respectively a) one, b) two and c) three layers are placed on top of the fixed layer. The global optimum structures are shown as insets with the colors of fixed atoms dimmed.}
	\label{fig:TiO2}
\end{figure}

Improved structure search methods allow for the determination of increasingly complex structures within an appealing time frame. To show that this is indeed the case for the method presented in this work, the method is applied to the anatase $\text{TiO}_2(001)\text{-}(1\times4)$ surface reconstruction and the complexity of the problem is increased by optimizing from one to three layers on top of a fixed layer as shown in the insets of \cref{fig:TiO2}. As for $\text{TiO}_2$ in the main text we compare to the well established evolutionary algorithm (EA) [L. B. Vilhelmsen and B. Hammer, J. Chem. Phys. \textbf{141}, 044711 (2014)] compared to which it is orders of magnitude faster and handles the increased complexity better. The method is also compared to the same EA for which machine learning is used in the form of clustering to improve the candidate generation step as in [K. H. S{\o}rensen, M. S. J{\o}rgensen, A. Bruix, and B. Hammer, J. Chem. Phys. \textbf{148}, 241734 (2018)]. This results is a significant improvement of the EA, but does not come close to the improvement achieved by GOFEE which uses machine learning to avoid the large number of first principles calculations traditionally spent on local relaxation.

\section*{Graphene edge structure}
\begin{figure}
	\centering
	\includegraphics[width=0.8\linewidth]{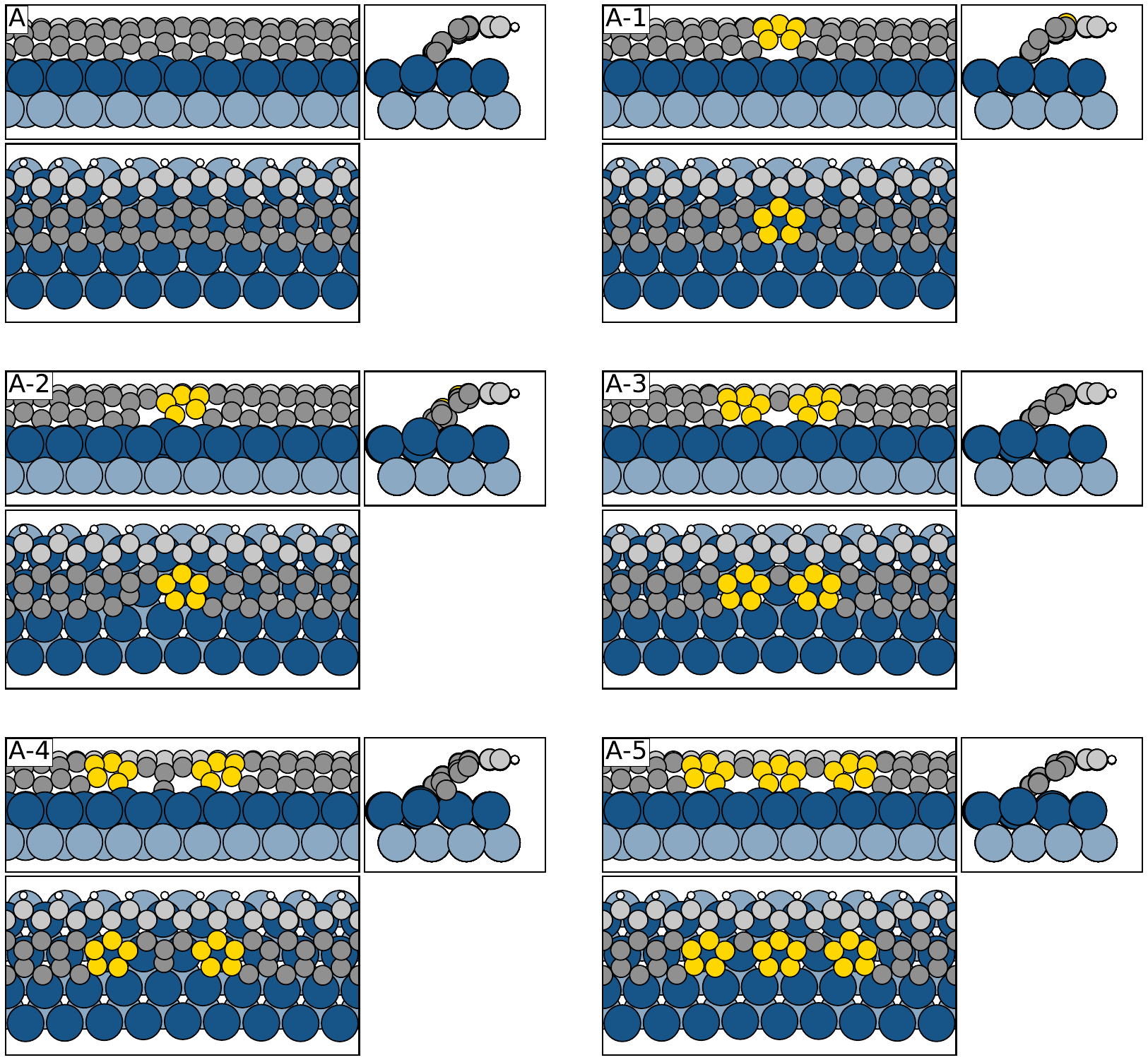}
	\caption{Top, front and side view of the graphene edge structures found using GOFEE in the cases of 0-5 carbon atoms removed from the perfect edge structure (A). In all cases where carbon atoms are removed, the deviations from the perfect edge (A) is centered at the position where the mismatch between the bonding carbon and iridium atoms are largest. Additionally all edges lacking carbon relative to (A) is seen to feature pentagonal rings as highlighted with yellow.}
	\label{fig:full_C}
\end{figure}

\begin{figure}
	\centering
	\includegraphics[width=0.8\linewidth]{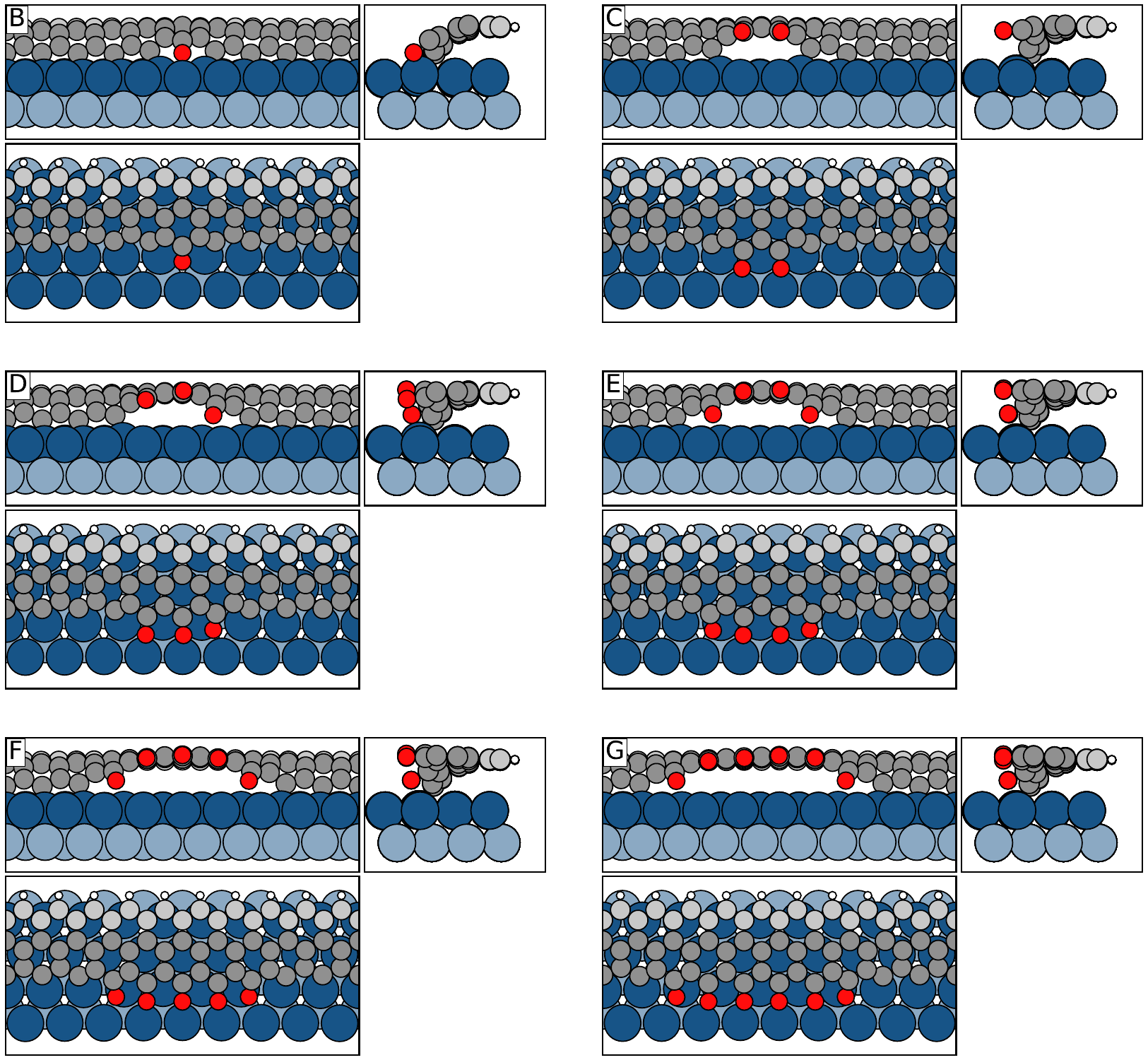}
	\caption{(B)-(D) Top, front and side view of the graphene edge structures found using GOFEE in the cases of adding 1-3 oxygen atoms to the perfect edge structure \cref{fig:full_C}A. (E)-(G) Structures found by extending the tendency seen in (B)-(D).}
	\label{fig:full_O}
\end{figure}

\begin{figure}
	\centering
	\includegraphics[width=0.8\linewidth]{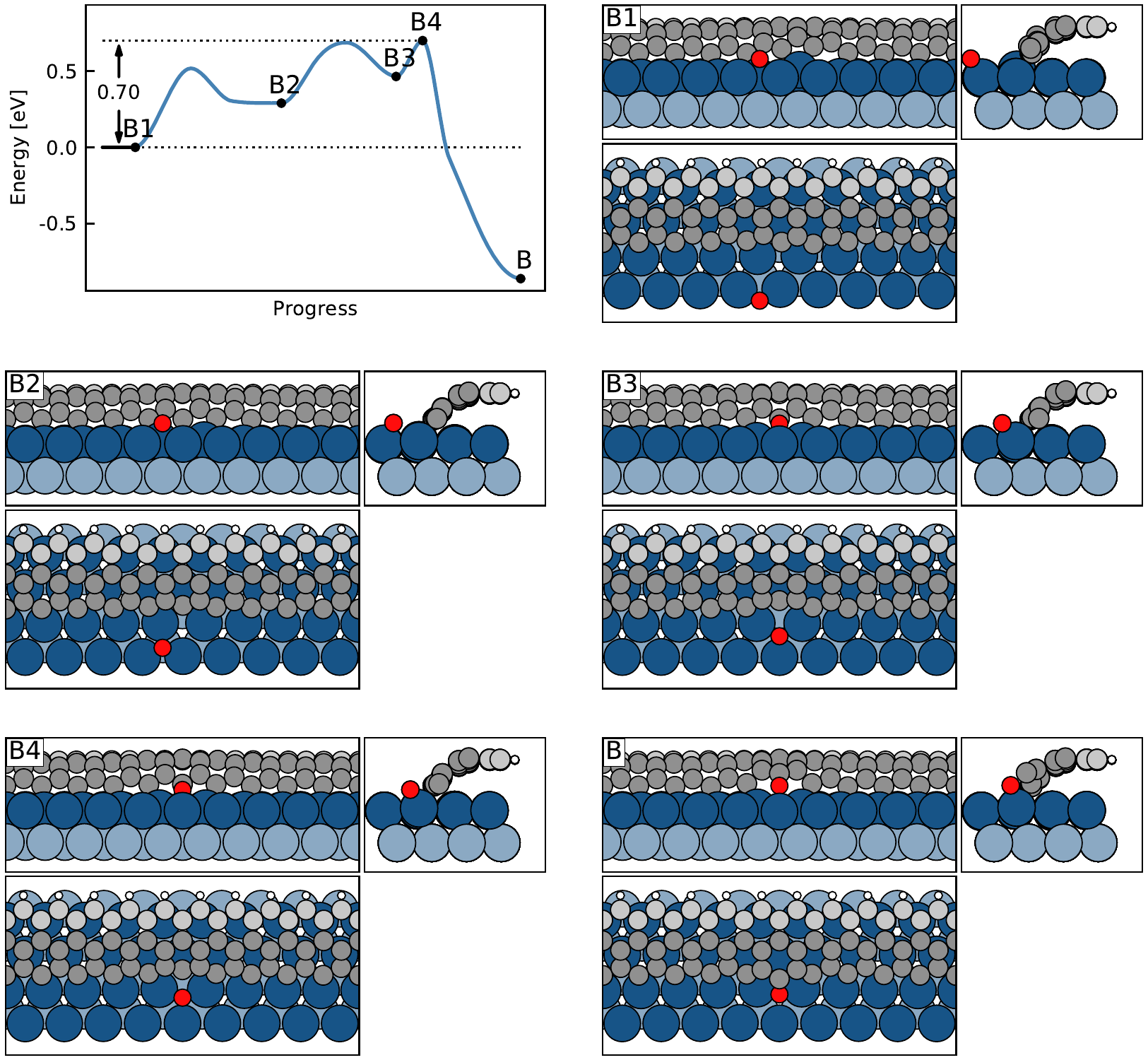}
	\caption{Lowest energy profile for the addition of a surface adsorbed oxygen atom to the perfect edge \cref{fig:full_C}A to form the edge structure \cref{fig:full_O}B. Top, front and side views are shown for selected structures on the pathway.}
	\label{fig:neb_ab}
\end{figure}

\begin{figure}
	\centering
	\includegraphics[width=0.8\linewidth]{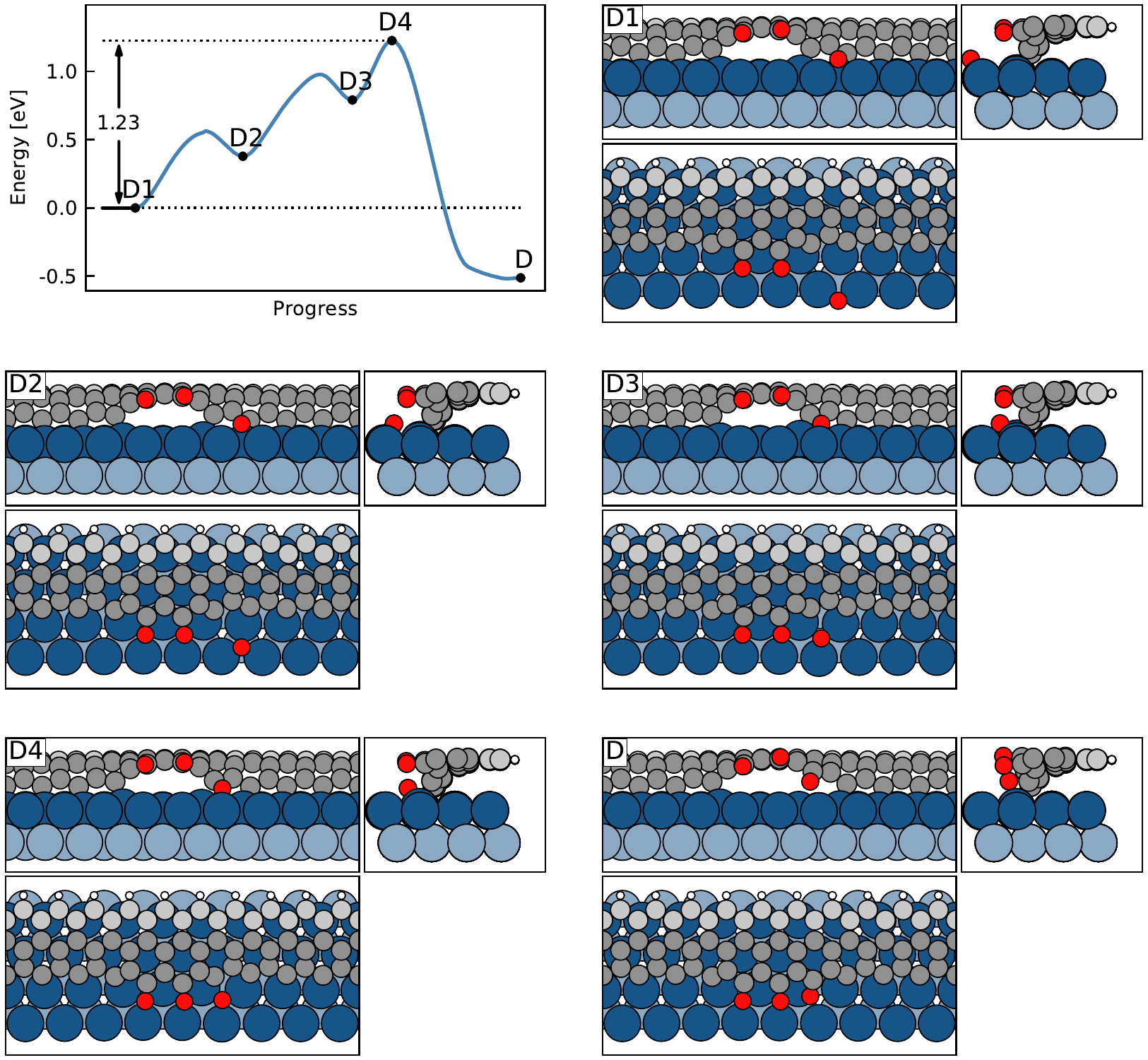}
	\caption{Lowest energy profile for the addition of a surface adsorbed oxygen atom to the edge structure \cref{fig:full_O}C to form the edge structure \cref{fig:full_O}D. Top, front and side views are shown for selected structures on the pathway.}
	\label{fig:neb_cd}
\end{figure}

\begin{figure}
	\centering
	\includegraphics[width=0.8\linewidth]{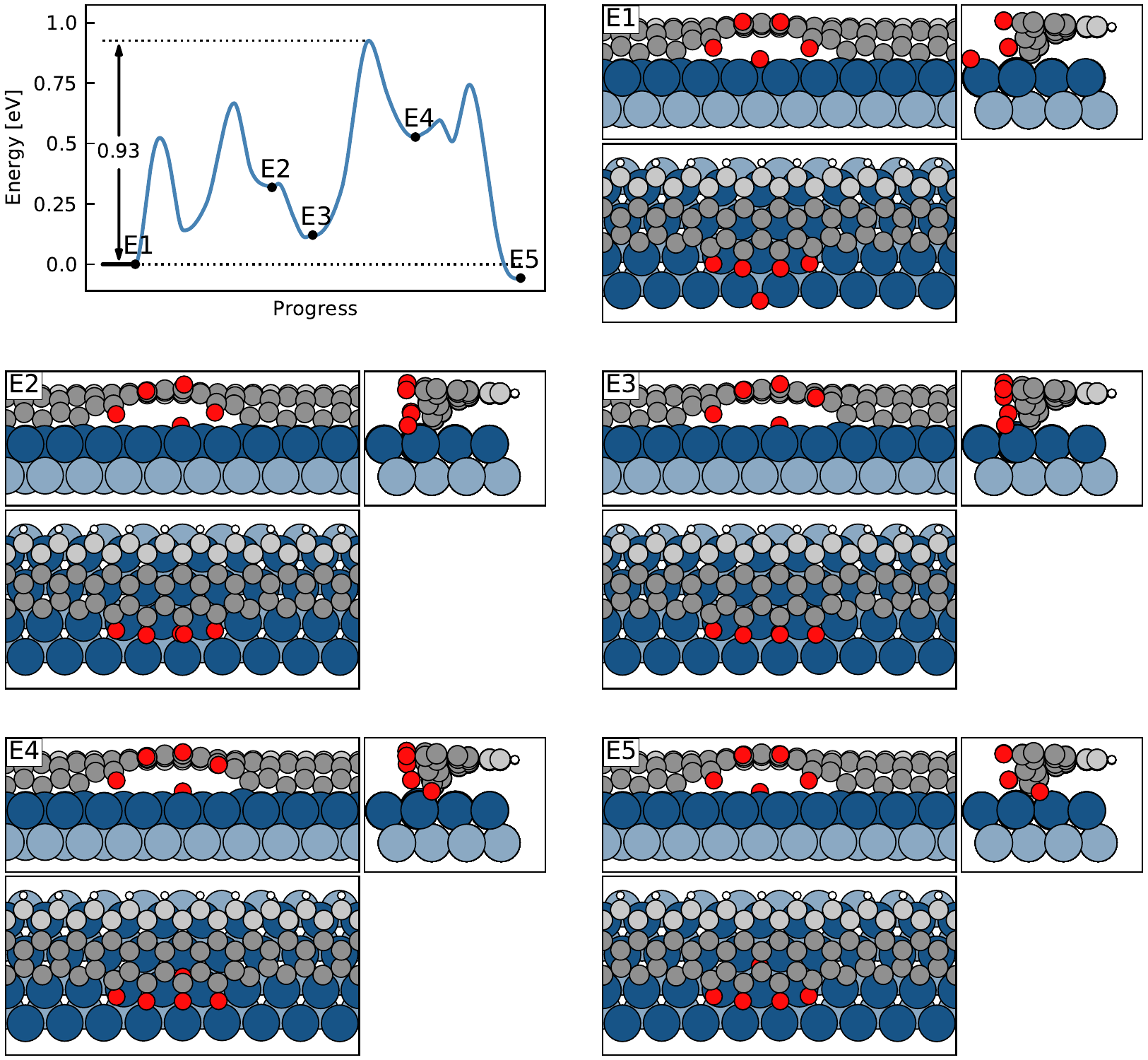}
	\caption{Lowest energy profile for the intercalation of a surface adsorbed oxygen atom below the open edge of the oxidized edge structure \cref{fig:full_O}E. It is notable that one of the edge oxygen atoms detaches from the surface during this process to open the edge further and accommodate the intercalation.}
	\label{fig:diff}
\end{figure}

Larger plots as well as top and side views of the graphene edge structures presented in the main article is shown in \cref{fig:full_C} and \cref{fig:full_O}. In addition the presented minimum energy profiles are shown in \cref{fig:neb_ab}, \cref{fig:neb_cd} and \cref{fig:diff} with snapshots of structures along the pathways.

\end{document}